\documentclass[10pt]{aastex}
\usepackage{bm}
\usepackage{lscape}
\usepackage[dvips]{epsfig}
\usepackage{amsmath}
\usepackage{amssymb}
\usepackage{amsthm}
\usepackage{array}
\usepackage{multirow}
\usepackage{graphicx}
\usepackage{amsfonts}
\usepackage{eucal}
\newcommand{\oq}{\textquotedblleft}
\newcommand{\cq}{\textquotedblright}
\newcommand{\cqb}{{\textquotedblright}~}
\newcommand{\ms}{$M_{\odot}$}
\newcommand{\msb}{$M_{\odot}$~}
\newcommand{\ct}{$^{13}$C}
\newcommand{\cd}{$^{12}$C}

\newcommand{\ctb}{$^{13}$C~}
\newcommand{\ctanb}{$^{13}$C($\alpha$,n)$^{16}$O~}

\newcommand{\neanb}{$^{22}$Ne($\alpha$,n)$^{25}$Mg~}

\title{\large$s$-Processing in AGB Stars Revisited. II. \\
Enhanced $^{13}$C Production Through MHD-Induced Mixing}

\author{O. Trippella, M. Busso, S. Palmerini, E. Maiorca}
\affil{Department of Physics, University of Perugia, and INFN, Section of Perugia, via A. Pascoli, I-06123 Perugia, Italy; oscar.trippella@fisica.unipg.it}

\author{M.C. Nucci}
\affil{Department of Mathematics and Informatics, University of Perugia, via Vanvitelli, I-06123 Perugia and INFN, Section of Perugia, via A. Pascoli, I-06123 Perugia, Italy.}

\begin{document}


\section*{Abstract}
Slow neutron captures are responsible for the production of about $50\%$ of elements heavier than iron, mainly, occurring during the asymptotic giant branch phase of low-mass stars ($1$ $\lesssim M$/\msb $\lesssim$ $3$), where the main neutron source is the \ctanb reaction. This last is activated from locally-produced \ct, formed by partial mixing of hydrogen into the He-rich layers. We present here the first attempt at describing a physical mechanism for the formation of the \ctb reservoir, studying the mass circulation induced by magnetic buoyancy and without adding new free parameters to those already involved in stellar modelling. Our approach represents the application, to the stellar layers relevant for $s$-processing, of recent exact, analytical 2D and 3D models for magneto-hydrodynamic processes at the base of convective envelopes in evolved stars in order to promote downflows of envelope material for mass conservation, during the occurrence of a dredge-up phenomenon. We find that the proton penetration is characterized by small concentrations, but extended over a large fractional mass of the He-layers, thus producing \ctb reservoirs of several $10^{-3}$ \ms. The ensuing \ct-enriched zone has an almost flat profile, while only a limited production of $^{14}$N occurs. In order to verify the effects of our new findings we show how the abundances of the main $s$-component nuclei can be accounted for in solar proportions and how our large \ct-reservoir allows us to solve a few so far unexplained features in the abundance distribution of post-AGB objects.

\newpage

\section{Introduction}
Most nuclei heavier than iron owe their existence to neutron ($n$) captures occurring in stars; about $50\%$ of them are produced in the final Asymptotic Giant Branch (AGB) phases of moderately massive stars \citep[$M \lesssim 8$ \ms, see][the highest efficiency being found between 1 and about 3 \ms]{IBE83}. There, the time scale for neutron captures is usually much longer than that for the $\beta^-$ decay of unstable nuclei along the $s$-process path, so that branching reactions are rarely encountered. For this reason the process is called the {\it slow} (or \oq $s$-\cq) process. The largest part of the nuclear phenomena controlling the abundances of $s$-elements from Sr to Pb (the \textit{main component} of the $s$-process) occurs in stars with masses from one to a few times solar \citep{KAE90}. Although about $90\%$ of galactic stars undergo these stages, the mechanisms through which effective neutron sources are made available have so far eluded our attempts at building quantitative, non-parametric models for their understanding \citep{BUS99}. 

We hope to have made here a substantial step forward in that direction, presenting a self-consistent model for the formation of the main neutron source for $s$-processing in the He-rich layers of Asymptotic Giant Branch (AGB) stars of low mass ($M \lesssim 3$ \ms). As described, e.g. in \citet{KAE11}, these stars undergo repeated instabilities of shell He-burning in thermal runaways (\oq Thermal Pulses\cq, hereafter TP), when the H-burning shell is switched off temporarily. The large energy release due to the 3$\alpha$ process being confined to thin stellar layers, cannot be transported by radiative diffusion; rather, convective energy transport becomes more efficient in this case. Hence an intermediate convective zone is formed in the He-rich region and all the materials external to the degenerate C-O core (sited below the helium shell) are forced to expand and cool. The innermost border of the envelope convection then extends inward beyond the molecular weight discontinuity left by H-burning, carrying freshly synthesized materials to the surface, in various mixing phenomena collectively indicated as the \oq Third Dredge-up\cq, or TDU \citep[see e.g.][]{BUS99,KAE11}.

In the phases subsequent to the development of the intermediate convective layer, some penetration of protons creates a reservoir of hydrogen in the He-rich layers. When the outer parts of the star re-contract and heat, H-burning is ignited again and the trapped protons are captured by the abundant \cd, inducing the chain $^{12}$C(p,$\gamma$)$^{13}$N($\beta^+\nu$)$^{13}$C. At this point the star presents a \ct-reservoir (hereafter \textit{pocket}) embedded in a He-rich environment, so that when the temperature approaches $(0.9$ $-$ $1.0)\times 10^8$ K, the reaction \ctanb is activated, releasing neutrons. This was actually recognised as being the most important source of neutrons in AGB stars \citep{GAL88}; it is so because, when the stellar mass is sufficiently low ($M \lesssim$ $3$ \ms), the second important neutron-source \neanb is only marginally activated (due to the moderate temperature, $T \lesssim 3\times 10^8$ K) in the convective TP that follows.

As mentioned, a physical model for the formation of the \ct-pocket, exempt from the introduction of further free parameterizations, has not been developed so far. Despite this,  we learned a lot from models allowing the \ct-pocket to vary,  and by fixing its extension from comparisons with the observations. However, we think that now a step forward can be done, using the knowledge acquired in recent years by the many studies on non-convective mixing \citep[see e.g.][]{BUS07,DEN11}. Our aim is also to provide physical tools to decide in the present debate on the \ct-reservoir extension. Roughly speaking, one sees today two alternatives: on one side there are models like those by \cite{GAL98,CRI09}, where the layers enriched in \ctb are assumed to be generated by some diffusion or incomplete mixing at the envelope border, and hence to be rather small (a few $10^{-4}$ \ms). On the other side, new computations have guessed that the \ct-reservoir might attain much larger extensions, on the basis of evidence coming from young Open Clusters \citep{MAI11}, from galactic chemical evolution \citep{MAI12}, from extragalactic abundances \citep{MCW13} and from the accurate analysis of isotopic anomalies in trace elements of presolar grains \citep{LIU14A,LIU14B,LIU15}. These issues were recently discussed in \citet{TRI14} and updated models by \citet{CRI15A} essentially confirmed such suggestions. In this recently-emerged view the mass of the \ct-rich layers covers a substantial extension of the He-rich zone, permitting Galactic evolution models to account for the whole main component, without the need of introducing new nuclear processes (sometimes called LEPP, or Light Element Primary Processes), as previously done. In any case, an intrinsic spread of $s$-process efficiencies is observed in real AGB stars \citep{BUS01} so that what is needed is a physical model capable 
to provide a range of different proton penetration efficiencies. 

Similar needs for deep mixing exist above the H-burning shell, for explaining the isotopic composition of light elements of Red Giant Branch (RGB) and AGB stars and for reproducing the isotopic admixture of intermediate elements (especially C, N, and O) in presolar grains of stellar origin \citep[see e.g.][]{CHA94,WAS95,NOL03}. Some years ago \citet{BUS07} suggested that magnetic stellar activity, known to be present in Low Mass Stars (LMS) from many observational constraints, might play a role, by inducing the buoyancy of materials lifted by magnetic pressure as a consequence of a dynamo mechanism. This buoyancy provides a physically-based scenario, being related to the key properties of a stellar plasma. Subsequently, \citet{NUC14} demonstrated that, under the special conditions holding in sub-convective zones of AGB phases, the full Magneto-Hydrodynamic (MHD) equations strongly simplify and can be solved analytically in an exact way. They guarantee a steady expansion of magnetized domains, confirming the hypothesis by \citet{BUS07}. A stable circulation can thus be established, with characteristics suitable to account for deep mixing.

In this paper we will show that the same explanation \citep{BUS07,NUC14} holds also for the formation of the neutron source \ctb in He-rich layers, thus providing for it the long-waited-for link to sound physical principles.

In section $2$ we verify whether the conditions identified by \citet{NUC14} as sufficient to induce a stable magnetic buoyancy process hold for the He-rich layers of an AGB star after a TP. Then in section $3$ we present the consequences of this for the establishment of p-rich downflows and we derive the shape expected for the ensuing proton profiles in the He-layers. Then, in section $4$ we compare the nucleosynthesis results obtained with our MHD-induced \ctb reservoir with observational constraints from the solar composition and from post-AGB stars. This comparison adds to the mentioned evidence already existing from the chemical evolution of the Galaxy, showing that our proposed model for the \ct-pocket works well. The formation in AGB stars of \ctb reservoirs that in most cases extend for a few $10^{-3}$ \ms, as early suggested by \citet{MAI11, MAI12} and \citet{TRI14}, seems now to be rooted in fundamental plasma processes of these advanced evolutionary stages.

\section{Buoyancy in the He-intershell}
\citet{NUC14} used initially a two-dimension geometry and then extended it to a $3$D computation of the buoyancy velocity, showing that in the radiative zones of an AGB star a rather general, exact, analytical solution is possible for the full MHD equations. The conditions sufficient for the existence of this simple, exact solution were shown to require the occurrence of three different facts: i) in the relevant zones the plasma must have a simple density distribution, of the form $\rho \propto r^k$, where $r$ is the distance from the stellar core (rotating as a rigid body) and $k$ is a negative number such that $k < -1$ (as an example we see from Figure 1 that in the layers we are dealing with $k = - 4.4$); ii) in the relevant layers a quasi-ideal MHD condition must hold\footnote{Here the term \oq quasi-ideal\cqb is used, as in \citet{NUC14}, to mean that at least the kinematical viscosity $\eta$, albeit small in itself, cannot really be neglected, as it remains much larger than the magnetic diffusivity, to provide large values of the magnetic Prandtl number.}. This means that dissipative terms of the MHD equations (i.e. the dynamical viscosity $\mu$, the magnetic diffusivity $\nu_m$, the coefficient of thermal conduction $\kappa_T$) must be small; iii) the \textit{kinematical} viscosity $\eta =\mu/\rho$ (with $\rho$ representing the density) must remain considerably larger than the magnetic diffusivity $\nu_m$, so that their ratio, also called \textit{magnetic Prandtl number $P_m$}, must be much larger than unity \citep[see][]{SPI62}. If these three rules are satisfied, when toroidal magnetic structures are generated in radiative layers by a dynamo process, they are subject to a \textit{necessary} buoyancy phenomenon induced by their extra magnetic pressure.
As shown in the mentioned work by \citet{NUC14}, the radial component of the outflow velocity and the toroidal component of the magnetic field can then be written as: 
\begin{equation}
\label{eq1}
v_r=\dfrac{dw(t)}{dt}r^{-(k+1)}
\end{equation}
\begin{equation}
B_{\varphi}=\Phi(\xi)r^{k+1}
\label{eq2}
\end{equation}
where
\begin{equation}
\xi=w(t)+r^{k+2}.
\label{eq3}
\end{equation}
Here $w$ and $\Phi$ are mathematically arbitrary functions, which need to be physically specified by boundary conditions. Note that the solutions allow for both time-dependent and constant magnetic fields, as it is expressed via the function $\Phi(\xi)$ that can be fixed arbitrarily. This is one of the advantages of the very general analytical solutions found by \citet{NUC14}. In particular, $\Phi$ can be for simplicity chosen to be a constant. Then the boundary conditions are such that $w(t)$ can be chosen arbitrarily. The simplest solution admitting buoyancy is:
\begin{equation}
w(t)=\Gamma t,
\label{eq4}
\end{equation}
where $\Gamma = v_p r_p^{k+1}$. Then we can write:
\begin{equation}
v_r= v_p \left(\frac{r_p}{r}\right)^{k+1} 
\label{eq5}
\end{equation}
and:
\begin{equation}
B_{\varphi} = B_{\varphi,p} \left(\frac{r}{r_p}\right)^{k+1},
\label{eq6}
\end{equation}
where $B_{\varphi,p}$ is constant. Here, with the suffix \oq $p$\cqb we indicate the values pertaining to the layer from which buoyancy starts (see later, and the fourth row in Table \ref{tab1}). In general we see that the form of the 
buoyancy velocity inducing mixing depends on radius as a power law with exponent: $-(k+1)$. 

\begin{figure}[t!]
\centering
\includegraphics[width=0.55\textwidth]{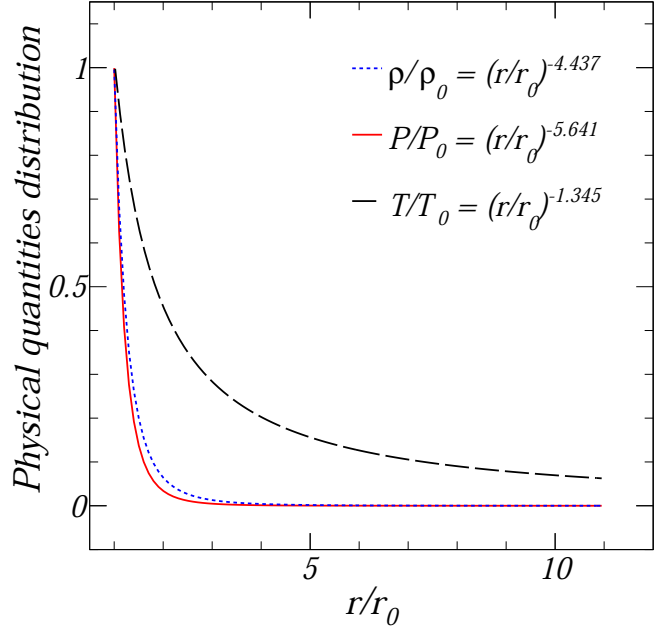}
\caption{\label{one}The pressure (${\rm P}$), density (${\rm \rho}$) and temperature (${\rm T}$) distributions as a function of radius in the \textit{intershell} region of a star with $1.5$ \msb and solar metallicity. The curves are power-laws with exponents indicated and regression coefficients $R^2$ always larger than $0.993$. The abscissa indicates the radius, and covers an interval in mass of about $0.02$ \ms, up to the inner border of the convective envelope. This is larger than any value of the \ct-pocket ever explored. The parameters shown are normalized to their values at the innermost point considered (layer \oq $0$\cq). The Figure demonstrates that the density varies as $r^k$, with $k= -4,437$, i.e. a value considerably smaller than $-1$, as it must be for the validity of the model.}
\end{figure}

Choosing as an example the stellar model discussed by \citet{BUS07}, for a $1.5$ \msb star of solar metallicity, one sees that at the development of a TDU episode, the dependencies of $P, T$, and $\rho$ on the radius have the form illustrated in Figure \ref{one} (we chose the $6^{th}$ TP for the sake of exemplifying).

We can see from Figure \ref{one} that the relation:
\begin{equation}
\rho(r)=\dfrac{\rho_p}{r_p^k}r^k 
\label{eq7}
\end{equation}
holds with a high accuracy (with a regression coefficient $R^2= 0.994$) and $k$ is approximately $-4.4$. Hence the first condition found by \citet{NUC14} for obtaining a solution suitable to guarantee stable mass transport is satisfied. Actually, it is satisfied even better than for H-rich layers, as the absolute value of $k$ is very large: magnetic buoyancy can therefore guarantee a velocity of transport for He-rich materials into the envelope which easily becomes very fast. This is important, in view of the short time interval available for mixing before the reignition of the H-burning shell (typically less than $1/3$ of the interpulse period, i.e. a few thousands years, reduced to about a hundred for the actual phase of TDU). 

One has now to verify that also the other two conditions hold. 

In order to do this, Table \ref{tab1} shows a typical physical situation in the He-rich layers at the same TDU episode of Figure \ref{one}, from the model discussed by \citet{BUS07} and \citet{NUC14}. 

Following the approaches by \citet{CHA51,PAR60,SPI62}, and \citet{NUC14}, the parameters of interest (dynamical viscosity and magnetic Prandtl number) are shown in the two last columns of Table \ref{tab1}, as given by the formulae:
\begin{equation}
\mu = \lambda c_s \rho
\label{eq8}
\end{equation}
\begin{equation}
P_m \simeq 2.6\times 10^{-5} T^4/n
\label{eq9}
\end{equation}
where $c_s$ is the sound speed ($c_s = \sqrt{\gamma(P/\rho)}$), $\gamma$ is the adiabatic exponent, $\lambda$ is the De Broglie wavelength for the He ions and $n$ is the number density of the plasma. The small values of $\mu$ and the values of $P_m$ significantly larger than unity over most of the He-rich layers certify that the same quasi-ideal MHD conditions found by \citet{NUC14} above the H-burning shell also hold below the convective envelope at TDU, at least over a layer reaching almost $0.01$ \ms; see Table \ref{tab1} for more details. This is a zone safely larger than any extension of the \ct-pocket explored so far \citep{MAI11,MAI12,TRI14,CRI15A}. All the conditions for a fast circulation induced by quasi-ideal MHD are therefore well satisfied.

Hence we can expect that magnetic buoyancy pushes rapidly matter from the He-rich layers (homogenized by the occurrence of a TP) to the envelope, thus inducing also a downflow of protons for maintaining mass conservation across the envelope. In the next section we shall now explore how this downflow can plausibly occur.

\section{Non-parametric proton downflows}
When matter is pushed up into the envelope from below, driven by the buoyancy of magnetic flux tubes, as outlined in section $2$, conservation of mass implies that downflows rich in protons are forced to occur. If \citep[as discussed by][]{NUC14}, magnetized structures are rapidly destroyed in the convective envelope by macroturbulence, these downflows will be no longer confined inside magnetic structures, but will spread across the whole upper boundary of the radiative layer. Let us now consider a unit volume at the interface of convective/radiative zones and a (polar) reference system in which the radial axis be oriented like the stellar radius. Let us also indicate by $\rho_d$ the mass per unit volume pushed downward and let $r_e$ indicate the position of the inner convective border measured in units of the stellar radius $R_{\ast}$ (which is about $400-500$ $R_{\odot}$ in the phases considered). In every AGB condition (where most of the stellar volume is occupied by the huge convective envelope) $r_e$ is a very small number (typically between $10^{-4}$ and $10^{-3}$: see e.g. Table \ref{tab1}). The density of envelope material injected into the He-layers will vary, in travelling toward \textit{smaller} values of the radius by an elementary displacement $-dr$, as:
\begin{equation}
\rho_d(r) - \rho_d(r-dr) = -\rho_{d} \alpha(-dr),
\label{eq10}
\end{equation}
where $\alpha$ is either a constant or a function of $r$ \citep[see e.g.][]{TRI14}. Let us assume here, for simplicity, that $\alpha$ is a constant. In this case one reaches a distribution with an exponential profile:
\begin{equation}
d \rho_{d}/\rho_d = + \alpha dr
\label{eq11}
\end{equation}
\begin{equation}
\rho_d(r)=\rho_{d,0} e^{-\alpha(r_e-r)}
\label{eq12}
\end{equation}
Here $\rho_{d,0}$ is the mass per unit volume crossing the envelope border in going down to the radiative layers. Quite obviously, as at any specific value of the radius we have a corresponding value of the mass, equation (\ref{eq11}) can also be expressed as a negative exponential in mass (with a different value of $\alpha$) as done e.g. by \citet{TRI14}.
We shall adopt $\rho_{d,0} \simeq \rho_e$, the total mass density at the border.

Now, what has always been considered as important for defining the efficiency of $s$-processing is the extension in \textit{mass} of the proton penetration. However, in our case (see the previous section) we have all the parameters for the buoyant part of the circulation expressed as a function of radius; it is therefore convenient to maintain here the same approach for coherence. If the layer polluted is sufficiently deep in radius one cannot neglect the stellar curvature (the connection between the depth in mass and the depth in radius will obviously depend on the local mass distribution in the relevant stellar layers). In general, the elementary volume will be $dV = 4\pi r^2 dr$, so that the mass entering the He-rich layers over a distance $dr$ is:
\begin{equation}
dM_d(r) = 4\pi r^2 \rho_e e^{-\alpha(r_e - r)} dr.
\label{eq13}
\end{equation}
If instead the extension of the penetration in radius is small, neglecting the curvature yields also for the total mass penetrated a simple exponential form. From what we know of the spread of $s$-process abundances in observed AGB stars, we expect that such small pockets may in fact exist, but also that their effects on the chemical evolution of the Galaxy be relatively small \citep{MAI12}.

We must notice that the flux tubes transport very little mass: they are almost empty, as the mass in them is concentrated in thin current sheets with a filling factor of about one hundredth of the tube section \citep[see e.g.][]{HIR92}. The tubes themselves represent concentrations of the magnetic field which cover a fraction of the surface $f_2$ that, at the base of convective envelopes, should be typically less than $1/1000$, as confirmed by the Sun \citep[see][]{NUC14}.
 
Assuming that the circulation rate induced by buoyancy, $\dot M$, is constant over the short time interval $\Delta t$ over which the TDU episode reaches its maximum downward extension ($50$ $-$ $100$ years), one has that the mass $M_{up}$ transported by buoyancy is:
\begin{equation}
M_{up} = \dot{M} \cdot \Delta t = 4\pi r_e^2\rho_e v_e f_1 f_2 \Delta t
\label{eq14}
\end{equation}
As we are in layers more internal than those discussed by \citet{NUC14}, characterized by higher densities, we expect $v_e$ to be substantially smaller than the value assumed by those authors ($100$ m/sec). For the sake of exemplifying we shall use in what follows $v_e=10$ m/sec, $f_1 = 1/100$, and $f_2 = 1/2000$. As we shall see, these assumptions are not critical.

Then the mass transported upward must equal the one ($M_{d}$) that is carried down by the penetration of envelope material. Its rate is obviously: 
\begin{equation}
\dot M_{d} = 4 \pi \rho_e r^2 e^{-\alpha(r_e - r)} v_d(r) 
\label{eq15}
\end{equation}
which yields to the formula in (\ref{eq13}):
\begin{equation}
dM_{d} = 4 \pi r^2 \rho_e e^{-\alpha(r_e-r)} dr.
\label{eq16}
\end{equation}
Integration of equation (\ref{eq16}) must be performed over the layers where the viscosity remains sufficiently small such that the treatment by \citet{NUC14} of an almost ideal MHD holds: from Table 1 this corresponds to a mass thickness of at least $0.004$ and $0.005$ \msb and maybe more. It is remarkable that this is almost exactly the pocket mass estimated by \citet{MAI12} from Galactic Evolution constraints. In a simple, recursive integration by parts, the presence of the term $r^2$ produces a polynomial expression of the second degree that multiplies the exponential function. Thus, with very simple algebra one obtains:
\begin{equation}
\Delta M^H_d \simeq 0.714 \frac{4 \pi \rho_E}{\alpha}\left\lbrace \left[r_e^2-\frac{2}{\alpha}r_e+\frac{2}{\alpha^2}\right] - \left[r_p^2-\frac{2}{\alpha}r_p+\frac{2}{\alpha^2}\right] e^{-\alpha(r_e-r_p)}\right\rbrace,
\label{eq17}
\end{equation}
Here 0.714 is the abundance of H assumed in the envelope, while $r_p$ is the value of the radius at the mass layer up to which protons have penetrated. From what we have discussed in section $2$, one can identify this parameter with the level from which the buoyant magnetic structures start (a layer where a rather small value for the dynamic viscosity and a large value for the magnetic Prandtl number must still be simultaneously found).

\begin{figure}[t!]
\centering
\includegraphics[width=0.55\textwidth]{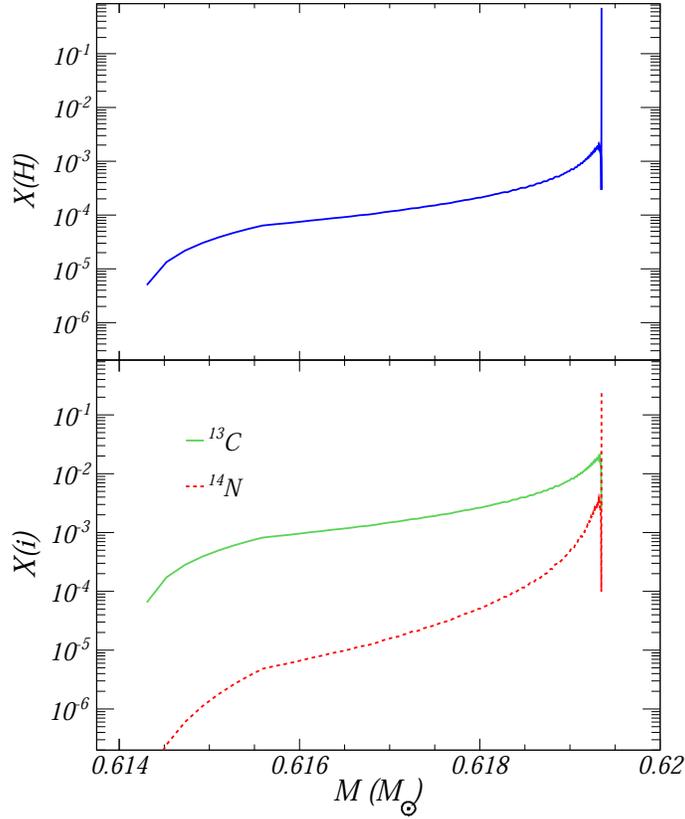}
\caption{\label{two}\textit{Upper panel}: Profile of proton penetration in the He-intershell layers during the third dredge-up phenomena of an AGB low-mass star. The hydrogen mass fraction $X(H)$ is plotted versus the stellar mass in units of \msb proceeding (left to right) from inner to outer regions of the star. The p penetration, due to the magnetic buoyancy, is around $5 \times 10^{-3}$ \ms. \textit{Lower panel}: Same as in the upper panel, but in the case of \ctb (green line) and $^{14}$N (red dashed curve). In this context, the so-called effective \ctb reservoir can be identify where the green line overcomes the red one. Assuming the proton profile described in the text and shown in the upper panel, no $^{14}$N-rich regions are expected as a direct consequence of the hydrogen nucleosynthesis occurring in the intershell region with the only exception of the first, but small and negligible, mesh on the right. (A color version of this figure is available in the online journal.)}
\end{figure}

Equating (\ref{eq14}) with (\ref{eq17}) yields the value of the unknown parameter $\alpha$. For the choices illustrated above for $v_e$, $f_1$ and $f_2$ one gets $\alpha \simeq 9820$. Its exact value (and hence the specific choice of the mentioned parameters) is not critical: the general outcome of this short discussion is simply that $\alpha$ must be very large for most situations, in which the proton penetration is forced by magnetic buoyancy.
This is not without surprising consequences. The form of the pocket enriched in envelope material that is created in the above model inside the He-rich layers is indeed rather peculiar. The abundance of H at the first mesh (the one anchored to the bottom of the convective envelope) is large; but as is illustrated in Figure \ref{two} the formula (\ref{eq17}) implies that it drops dramatically describing almost a discontinuity. Then it rises again for a very short time in the region where the exponential term is not too large, and reaches finally an almost horizontal plateau. 

Notice that the combined effects of a low viscosity down to very deep layers and of a rather inefficient mass transport by magnetic flux tubes reflects itself into a deep penetration of a very small amount of protons, so that when H burning restarts the protons themselves are captured almost exclusively by $^{12}$C, producing \ctb but very little $^{14}$N. This is so with exclusion of the first very thin layer at the base of the envelope, where the $^{14}$N abundance must be larger than that of $^{13}$C. This is however a region so small that has essentially no effects on the ensuing nucleosynthesis. This resulting composition profile is shown in Figure \ref{two} and is very different from the typical outcome of an exponential pocket, without the \oq minus\cqb sign and without the $r^2$ term of equation (\ref{eq17}) \citep[see e.g. results by][for a comparison of the two scenarios]{GAL98}.

The \ct-pocket thus obtained has some remarkable features recently invoked in the literature as required by various observational constraints. It is sufficiently extended to account for the chemical evolution of $s$-process elements, without requiring any ad-hoc n-capture process like the so-called \oq solar LEPP\cq, introduced by \citet{TRA04} \citep[after a few years of study on this topic, starting with][we can now be confident that this process does not occur in Nature]{MAI11}. Moreover, our pocket has a very flat abundance of \ct: this fullfils the requirements recently imposed by the isotopic composition of $s$-elements in presolar SiC grains of AGB origins \citep{LIU14A,LIU14B,LIU15}. 

\section{Comparison with observations}
In this section we shall compare the results of $s$-processing \citep[computed as described in][]{TRI14} with the solar distribution, in order to verify whether the new profiles of \ctb and $^{14}$N remain suitable to account for it. Subsequently we shall consider the composition of post-AGB stars. For them, explaining the total abundance ratio of $s$-elements to Carbon ($s$/C) has been so far impossible for parametric nucleosynthesis models \citep{PER12,DES12,DES14}.

\subsection{Reproducing solar abundances}
As a first application of our non-parametric, MHD-induced \ctb production in the He layers of an AGB star, we need to verify that the solar distribution of $s$-process elements can be reproduced using the new profiles of $^{13}$C and $^{14}$N. In so doing, we must remain coherent with the assumptions discussed in \citet{TRI14}, e.g. require that AGB stars undergo a limited number of thermal instabilities, in line with the observations of moderate magnitudes for C-stars \citep{GUA06,GUA13}.

As usual, while the quantitative, global solar abundance distribution will derive from the chemical evolution of the Galaxy, we expect that, with a suitable scaling factor, this distribution be mimicked by AGB models in the range $1.5$ $-$ $3$ \ms, computed at a suitable metallicity. Due to our large extension of the pocket, similar to those by \citet{MAI12} and \citet{TRI14}, a relatively high value is expected for this metallicity.

We also need as a minimum requirement that the new models produce fits to solar abundances with at least the same level of accuracy as those obtained previously with simple, parametric and exponential H-profiles. This would already be a good point as it would mean reproducing the solar system constraints without any tuning of parameters; but of course we aim at much more; some of the extra results that our model can produce will be illustrated in section 4.2.

\begin{figure}[t!]
\centering
\includegraphics[width=0.55\textwidth]{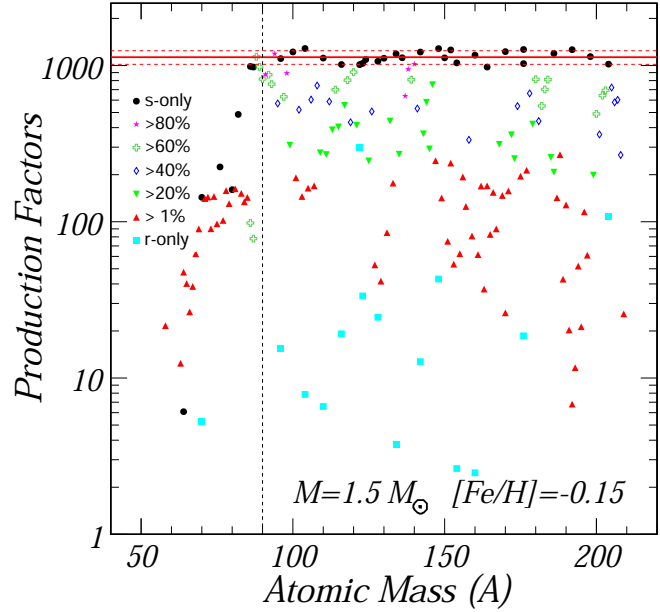}
\caption{\label{three}Production factors with respect to initial abundances of $s$-elements in the He-rich regions. The solar distribution is well reproduced adopting a stellar model of $1.5$ \msb and $[Fe/H]=-0.15$ that experiences a limited number (in the specific case, nine) of thermal pulses. The \ct-pocket, suggested and described in the text (see Figure \ref{two}), has been used for calculations with post-process nucleosynthesis code NEWTON \citep{TRI14}. In order to guide the eye, the red full line identifies the mean value of $s$-only nuclei that are not affected by branchings, and it is also very similar to $^{150}$Sm overabundance. The two red horizontal dashed lines represents the standard $10\%$ of fiducial interval including both model and nuclear uncertainties. Colours and symbols refer to different $s$-process contributions, as shown in the inner legend. (A color version of this figure is available in the online journal.)}
\end{figure}

In \citet{TRI14}, we computed the enhancement factors of $s$-process elements with respect to the corresponding initial composition for LMS models in the range between $1.5$ and $3.0$ \ms. The outcomes of each model were then weighted using the Salpeter's initial mass function (or IMF). The results of the averaging procedure are very similar to the calculations of the individual model at $1.5$ \ms, because the IMF favours lower masses in the weighting operation. For that reason, in this paper we performed calculations (through our post-process code Nucleosynthesis of Elements With Transfer of Neutrons, or NEWTON) limiting ourselves to the case of a $1.5$ \msb star, as shown in Figure \ref{three}. The assumptions for the set of nuclear cross sections and reaction rates are exactly the same as in \citet{TRI14}, to which the reader can refer for details. We simply recall that most n-capture cross sections are from the KADoNIS database, versions $0.3$ and $1.0$; $\beta$-decays are mostly from \citet{TAK87} and the solar abundances are from \citet{LOD09}.

Figure \ref{three} shows the production factors in the He-rich region at the last computed thermal pulse, for the nuclei in the atomic mass range $70 \leq A \leq 210$, with respect to their initial abundances. For the sake of simplicity, we don't use any normalization, but represent the absolute values of overabundances for neutron-rich elements. To guide the eye, the red line corresponds to the mean overabundance of nuclei that are produced only by the $s$-process, the so-called $s$-only nuclei. In particular the isotopes not undergoing reaction branchings are adopted in computing the average. This overabundance is of the order of one-thousand (in the specific case shown it is about $1120$) and is very similar to the production factor of $^{150}$Sm, often used as a reference in the literature on $s$-processing. The other two red dashed lines define a fiducial interval of $\pm10\%$, i.e. the typical uncertainty in the input parameters, taking into account both nuclear and observational effects. The $s$-only nuclei between $86 \leq A \leq 204$, represented by black full circles, are all bracketed by the two fiducial lines, thus showing roughly constant production factors, suitable to reproduce the solar distribution. This is so for the so-called \oq main component\cq of the $s$-process, which in our computations includes nuclei from strontium to bismuth. These results are very close (within the confidence level of a few percent) to those of case B in \citet{TRI14}. Also the metallicity choice is roughly the same, around [Fe/H]$=-0.15$. In order to distinguish the contributions coming from the $s$-process to different heavy nuclei, these are represented in Figure \ref{three} by using markers of various colors and symbols.

As mentioned, the solar distribution of Figure \ref{three} is built through a number of pulses small enough to fulfil the requirements coming from the recent infrared analysis of AGB luminosities \citep{GUA06,GUA08,GUA13}. For the specific model shown here (the one for a $1.5$ \msb star with $[Fe/H]=-0.15$) this number is $9$ \citep[about $3-4$ pulses more than reported in the online FRUITY repository according to suggestions by][]{TRI14}, and the luminosity is about $10^4$ $L_{\odot}$. In fact, recent stellar evolution models \citep{CRI11} state that the combination of a lower number of convective instabilities from the He-shell, with a larger efficiency of TDU episodes guaranteed by the adoption of new opacities, implies that the total amount of processed matter is similar to the one assumed previously by older parametric models achieving higher luminosities.

Concerning the $s$-process main component, this work confirms results by \citet{TRI14}, in saying that it starts at $^{86}$Sr and $^{87}$Sr. These two nuclei are not usually included in the main component, which is often assumed to start at $A=90$ \citep{KAE11}.
We recall that the weak component of the $s$-process in massive stars has very few (if any) constraints on which to base its computations: essentially, it must provide what the main component cannot do. In this respect our results imply modifications also to the weak component and a reanalysis of it seems now rather urgent.
There is a close agreement between our predictions for the solar main component and those by \citet{TRI14}; this means that the different profiles of \ctb and $^{14}$N are not crucial for the solar distribution. It is indeed well known since many years that the production of $s$-elements in AGB stars in proportions suitable to fit the solar distribution is essentially controlled, for the same choices of the nuclear parameters and of the stellar model, only by the number of neutrons captured by heavy seeds at each interpulse-pulse cycle \citep[$n_c$, see e.g.][for details]{BUS95,BUS01}. The average value of $n_c$ we find is around $16$, essentially the same as found in \citet{MAI12} and in \citet{TRI14} for the same extension of the pocket. We notice that the {\it total} number of neutrons here is considerably smaller than in \citet{TRI14} and in \citet{MAI12}; however, the different analytical formula for the proton profile (see equation 17) guarantees that $^{14}$N remains always very low (see Figure \ref{two}) and essentially does not compete with iron for neutron captures. The number of neutrons captured by iron can for this reason remain the same.
 
We also notice that the scarcity of $^{14}$N would lead to a very low production of $^{19}$F from the material of the pocket, once this is mixed inside the next pulse. $^{19}$F production is limited here to the material mixed in the pulse from regions above the \ctb pocket (where the H-burning ashes are rich in $^{14}$N). A reanalysis of F abundances in AGB stars might be therefore a crucial test to prove or disprove our model.

\subsection{Reproducing other crucial observational constraints}
As the solar distribution of $s$-elements is mainly controlled by the number of nuclei going to iron seeds, it does not constrain the \ct-profile if not for the fact that, in order to avoid the need of extra nuclear processes like LEPP, the neutron exposures must be very efficient, so that the average metallicity of the stars most efficiently contributing to the solar distribution must be rather high (and rather constant in time), as typical of the long evolutionary stages of the Galactic disc, during which the age-metallicity relation is essentially flat \citep[see][for a detailed discussion]{MAI12}. 

However, complementary indications that come from different astrophysical sites exist. They allow us to verify whether the hydrogen profile induced by MHD mixing is really capable of accounting for a larger number of constraints than in previous parametric choices. One such constraint lays in the recent suggestion \citep{LIU15} that extended reservoirs of \ct, characterized by a rather flat distribution, might be required to explain the isotopic and elemental admixture of $s$-elements in presolar SiC grains. However, the model discussed here remains oxygen-rich during the whole of its evolution. In order to verify the above suggestions we shall therefore present soon new calculations for higher masses (reaching C/O $\ge$ 1 even at high metallicities).

In any case we know that high C/O ratios are easily reached at low metallicity even for rather low masses. We can exploit this fact to compute neutron captures in a model star with conditions typical of some low-metallicity, post-AGB objects \citep{DES14}. Post-AGB stars can play a crucial role in the understanding of $s$-process nucleosynthesis. They represent the final evolutionary phase of low and intermediate mass stars, on their way from the Hayashi track to the planetary nebula stages. The time scales for the post-AGB phase predicted by stellar models are in range $10$ to $10^5$ yr, inversely depending on the core mass. Crucial in determining this duration are the treatment of current mass loss and its history along the AGB. Since this evolutionary phase is fast, post-AGB stars are rare; yet they are also intrinsically bright. Hence, there is a rather vast sample of data and information on them, both for the central stars and their circumstellar material. Comparing our model predictions with the observational data gives rise to relevant constraints about the absolute $s$-process enrichment with respect to carbon. 

In the present work, we focus our attention on a specific example: the source $J00441.04$-$732136.4$ (hereafter $J00441.04$), for which an accurate spectral analysis was performed by \citet{DES12} and \citet{DES14} (other examples of post-AGB stars will be investigated in a forthcoming paper).
 
$J00441.04$ is a peculiar stellar object: with an initial mass and metallicity of $1.3$ \msb and $Z=0.001$, respectively, this star is the only one in the Small Magellanic Cloud showing the SiC feature at $21$ $\mu m$; it also shows one of the highest overabundances of $s$-process elements, with a moderate C-enhancement. We then computed the nucleosynthesis ensuing from a \ct-pocket like the one described in section $3$ for the case of a $1.3$ \msb star with [Fe/H] = $-1.4$. This offers us an occasion to check our predictions against observations for both the $s$-process distribution and the total $s$/C abundance ratio. This last value has remained so far unexplained.

\begin{figure}[t!]
\centering
\includegraphics[width=0.55\textwidth]{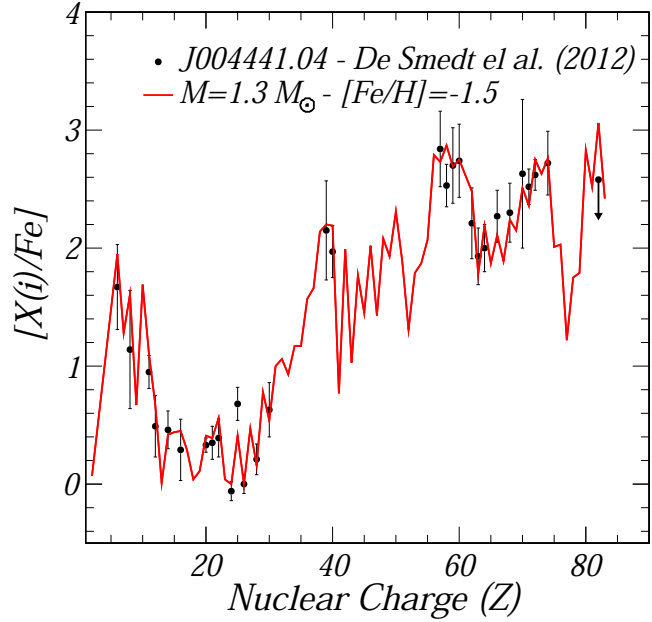}
\caption{\label{four}Comparison between model predictions (red line) and the $J004441.04$-$732136.4$ post-AGB abundance ratios ([X(i)/Fe]) identified by full black points and their corresponding error bars \citep{DES12,DES14}. A selection of elements, with nuclear charge in the range $2 \leq Z \leq 83$, is shown. Since the [Pb/Fe] data represents only an upper limit, it is identified by a down arrow. Our calculation does not show the enhancement factor of $\alpha$-element. (A color version of this figure is available in the online journal.)}
\end{figure}

In Figure \ref{four} we show our results for the [X(i)/Fe] ratios, as obtained by applying our post-process to a suitable AGB stellar model from \citet{STR03}. As the typical $\alpha$-enhancement for oxygen is usually larger than for other $\alpha$-rich elements, we assumed initial abundance values increased by $+0.6$ (O) and $+0.4$ (Mg, Si, S, Ca, Ti), respectively. 

In Figure \ref{four} our predicted abundances are represented by the red line and the observational data by the black points with the corresponding indication of the error bar. AGB nucleosytnthesis models based on small \ct-pockets can reproduce the average abundances of $s$-elements only after many thermal pulses, which means also obtaining necessarily a huge C abundance, as confirmed in \citet{DES12,DES14}. However, this is at odds with observational data. For the same reason, also the C/O ratio was not reproduced so far. These problems are not met by our model, whose fit is indeed very good for both light and heavy elements, within the uncertainties. This is an immediate consequence of our extended \ct-pocket, because it implies always a high $s$/C ratio in the material dredged up from the He-intershell. It was shown that the previous models can get close to the data only by assuming the existence of a late thermal pulse with a deep dredge-up and a small dilution with the convective envelope \citep[calculations were done with the STAREVOL code by][]{SIE07}. However, again the C/O ratio is not reproduced. We consider the automatic solution of all these problems as a striking evidence in favor of our model, in which $s$-elements are very effectively produced in few thermal pulses. We recall that this does not descend from any ad-hoc assumption and derives rather simply from our MHD-driven mixing mechanism.

A remaining problem for us, (which is actually met by any attempt with $s$-process codes) is the strangely moderate abundance observed for Pb, unexpected at this metallicity. Actually, it is an intrinsic nuclear property of $s$-processing to shift the abundances toward the magic nuclei at $N = 126$ at low metallicity, where the number of neutrons per iron seed becomes very large \citep{GAL98}. To our knowledge, this is an intrinsic nuclear property and is not model dependent; it leads us to suppose that something is wrong either with the measurement itself, or with the atomic physics parameters adopted for Pb. It has in any case to be noted that the discrepancy we find with the measured data (around $0.6$ dex) is significantly lower than in any other attempt so far made in the literature. Anyhow this is an important point to remember: if uncertainties in atomic physics parameters will be proven to be small and the moderate Pb observations will be confirmed, something crucial will require to be revised, both in this and in other models for the $s$-process.

Except for the case of Pb, only for manganese (Mn) our prediction is marginally outside the original error bar.

We can conclude that post-AGB stars can provide a striking support to our model of $s$-process nucleosynthesis where the \ct-reservoir is formed as a consequence of magnetic buoyancy. Our almost flat pocket of about $5\times10^{-3}$ \msb seems in fact to reproduce rather well the [X(i)/Fe] abundances of the $J00441.04$ source, together with its C/O ratio, without invoking free parameterizations.

\section{Conclusions}
In this paper we have applied the formalism developed by \citet{NUC14} to the mixing process driving, at each TDU episode in a thermally-pulsing AGB star, the formation of the proton reservoir in the He shell that subsequently induces the production of \ctb and the neutron release for $s$-processing, through the \ctanb reaction. We argue that magnetic fields generated in the He-rich zones, with a dynamo process similar to the solar one, can induce the buoyancy of magnetized parcels of matter to the envelope. There, conservation of mass and the disruption of magnetic flux tubes by stellar turbulence force some material with the envelope composition to be pushed downward beyond the He-H discontinuity. A knowledge of the magnetized transport then yields information on the form and extension of the proton reservoir and can serve to overcome the free parameterizations that have hampered our understanding of $s$-processing in stars over the last thirty years.

This work was motivated by three main reasons: i) recent revisions on deep mixing in stars, and in particular the suggestion that magnetic buoyancy can have a role in driving it \citep[see][]{BUS07,NOR08,NUC14}, gives us the opportunity, for the first time in many years of attempts, to base the formation of the \ct-pocket on \oq first principles\cq (albeit with a number of remaining limitations); ii) the observational evidence discussed in the text reveals that, while the galactic chemical evolution of $s$-elements requires that their production be dominated by stars having large reservoirs of \ct, observations of specific AGB stars show sometimes small abundances, suggesting an intrinsic spread of efficiencies \citep[see][]{BUS01,ABI02,ABI03}; something that appeared so far difficult to account by a non-parametric model and that motivated in the past the adoption of parametrized approaches; iii) recent constraints from post-AGB stars and from the analysis of $s$-element isotopes in presolar grains, when combined, call for a \ctb distribution that must be contemporary more extended and flatter than the commonly-used exponential behaviour.

In fact, the proton-abundance profile in our reservoir is exponential only per unit volume. The total hydrogen mass penetrated through the whole envelope border down to a depth $\Delta r$, once multiplied by the volume, keeps a memory of the stellar curvature. In particular, for large values of $\alpha$ equation (\ref{eq17}) can be rewritten as: $\Delta M \simeq k r_e^2{(1 - x^2 e^{-\alpha r_e^2 \Delta x})}$, where $r_e$ is the radius at the envelope border, and $x=r/r_e$ is a non-dimensional fractional radius. The parameter $\alpha$ is not free, as in most attempts so far, but descends directly from the mass conservation at the envelope border. Its value, for every reasonable choice of the physical conditions is very large (about $10^4$ in our calculations). The resulting distribution of \ct-nuclei after H-burning remains considerably flat. As mentioned above, that feature was recently suggested as a pre-requisite for $s$-processing to account for the heavy neutron-rich isotope abundances measured in presolar SiC grains of AGB origins \citep[][]{LIU14A,LIU14B,LIU15}. In a forthcoming dedicated paper we shall present a detailed application of our physical models to the interpretation of SiC grain compositions.

The extension of our proton reservoir is very close to the one suggested by \citet{MAI12} and by \citet{TRI14}. Although the slope is different, as mentioned, and the number of mixed protons is smaller, the very low abundance of $^{14}$N we produce makes the effects of neutron poisons less important, so that the effective number of neutrons captured by heavy seeds is essentially the same here and in the two mentioned works: we consider therefore the model discussed here as a generalization of those works and as a sound physical basis for the scenario that our group has depicted in recent years, suggesting that the extension of the \ct-pocket can often be much larger than assumed in the nineties.

This result is critical, as it allows us to get rid of other ad-hoc nuclear processes, like the so called \oq Solar LEPP\cqb previously sometimes invoked for explaining the chemical evolution of galaxies. We can now be confident that this process does not exist. We recall that, on this point, results compatible to ours were recently obtained, on completely different grounds, by \citet{CRI15A,CRI15B}.

We notice however that our model naturally allows also for the existence of AGB stars with more limited pocket extensions. Only galactic chemical evolution evidences tell us that they should be a minority, but in principle our equations (2) and (3) can link the buoyancy velocity to the magnetic field intensity through the function $w(t)$, which is arbitrary in the mathematical solution and therefore is compatible with several different physical situations. The values we used for the  rising velocity at the envelope border and for the filling factor of magnetic zones are aimed at giving typical average conditions, but different choices are in principle possible. In particular smaller filling factors and/or smaller velocities should be linked to smaller values of the magnetic fields in the stellar layers below the envelope. After this paper was submitted, we became aware of a recent important work from the Kepler space-borne observatory, which reveals that the existence of magnetic fields in the cores of evolved stars is very common; it also indicates that these fields are dispersed over a wide interval \citep{FUL15}. While the average values are in the range expected for the formation of a large \ct-pocket, the tail at lower $B$ values might induce (less frequently) the formation of smaller pockets, so that the puzzle of the above mentioned observed spread in the $s$-process efficiency can be simply explained by the observed spread in the intensity of magnetic fields in the stellar internal layers. We consider therefore the Kepler data as an important support to our scenario. 

The formation of the \ct-pocket described here represents therefore only the most common occurrence in a wider range of possibilities. As a remarkable consequence of the large number of neutrons that goes to Fe at every cycle in our \oq typical\cq pocket, our estimate of the $s$/C ratio in the material dredged-up from the He-shell to the envelope is significantly larger than in models discussed by e.g. \citet{BUS99} and \citet{BIS14}. We have shown in section $4.2$ that this makes AGB nucleosynthesis compatible with the abundances measured on C-rich post-AGB stars at low metallicity, an achievement impossible for the quoted parametric models \citep{DES12,DES14}. As mentioned in the text, a remaining problem is the abundance of Pb, which is not explained by any $s$-process model, including ours. The limited number of pulses that one now needs to form the $s$-process distribution allows us to account for the observational constraints on the magnitudes of $s$-process rich AGB stars \citep{GUA06}: this is another crucial agreement with observations.

Summing up, we think we have shown that if the \ct-pockets of AGB stars are formed thanks to a forced process induced by a dynamo mechanism, they have the characteristics suitable to account {\it in a single scenario}: i) for the solar distribution; ii) for the known constraints on AGB magnitudes; iii) for the abundances of some post-AGB stars; iv) for the existence of a spread in observed $s$-process abundances. To our knowledge, it is the first time that a physical model is shown to account for all these constraints without the need to accurately fine-tuning free parameters.

\section{Acknowledgements}
We are indebted to an anonymous referee for useful suggestions. This work was partially supported by the Department of Physics and Geology of the University of Perugia, under grant named \oq From Rocks to Stars\cq, and by INFN, through its Group 3 (Nuclear Physics and Astrophysics). O.T. is grateful to both these organizations for post-doc contracts.

\begin{table}[t!]
\centering
\rotatebox{90}{
\begin{tabular}{cccccccc}
\hline
\hline
$\Delta M$ (\ms) & Radius/R$_{\ast}$ & Mass (\ms) & $P (dyn/cm^2)$ & $T (K)$ & $\rho (g/cm^3)$& $\mu$ & $P_m$ \\
\hline
$0.000$ & $8.25\times10^{-4}$ & $0.61934$ & $3.34\times10^{11}$ & $2.94\times10^{6}$ & $9.83\times10^{-4}$ & $4.0\times10^{-6}$ & $4.7$ \\
$0.001$ & $2.89\times10^{-4}$ & $0.61835$ & $2.70\times10^{14}$ & $1.68\times10^{7}$ & $0.39$ & $4.3\times10^{-4}$ & $12.8$ \\
$0.004$ & $1.37\times10^{-4}$ & $0.61534$ & $2.80\times10^{16}$ & $4.29\times10^{7}$ & $7.81$ & $0.016$ & $27.0$ \\
$0.005$ & $1.09\times10^{-4}$ & $0.61434$ & $8.47\times10^{16}$ & $5.58\times10^{7}$ & $18.58$ & $0.041$ & $32.5$ \\
$0.010$ & $5.78\times10^{-5}$ & $0.60434$ & $1.82\times10^{18}$ & $1.12\times10^{8}$ & $222.18$ & $0.440$ & $52.3$ \\
\hline
\end{tabular}}
\caption{\label{tab1}Models parameters relevant for the calculations presented in sections $2$ and $3$ at various distances in mass ($\Delta M$) below TDU. Data refer to the $6^{th}$ TDU episode for the model star of $M = 1.5$ \ms, $Z = Z_{\odot}$ as discussed in \citet{BUS07}.}
\end{table}

\bibliographystyle{plainnat} 
\bibliography{trippella2015arxiv}

\end{document}